# High-Order Corrections to the Lipatov Asymptotics in the φ⁴ Theory

## D. A. Lobaskin and I. M. Suslov*

*Kapitza Institute of Physical Problems, Russian Academy of Sciences, Moscow, 117334 Russia*
*e-mail: suslov@kapitza.ras.ru

**Abstract**—High orders in perturbation theory can be calculated by the Lipatov method [1]. For most field theories, the Lipatov asymptotics has the functional form $ca^N \Gamma(N + b)$ ($N$ is the order of perturbation theory); relative corrections to this asymptotics have the form of a power series in $1/N$. The coefficients of high order terms of this series can be calculated using a procedure analogous to the Lipatov approach and are determined by the second instanton in the considered field theory. These coefficients are calculated quantitatively for the $n$-component $\varphi^4$ theory under the assumption that the second instanton is (i) a combination of elementary instantons and (ii) a spherically asymmetric localized function. The technique of two-instanton calculations as well as the method for integrating over rotations of an asymmetric instanton in the coordinate state are developed.

## 1. INTRODUCTION AND MAIN RESULTS

According to Lipatov [1], high orders of perturbation theory are determined by saddle-point configurations (instantons) of the corresponding functional integrals. A typical asymptotics for coefficients $Z_N$ in the expansion of any quantity $Z(g)$ in the coupling constant $g$,

$$Z(g) = \sum_{N=0}^{\infty} Z_N g^N, \tag{1}$$

has the form

$$Z_N = cS_0^{-N} \Gamma(N+b), \quad N \longrightarrow \infty, \tag{2}$$

where $S_0$ is the instanton action and $b$ and $c$ are certain constants. The corrections to the Lipatov asymptotics (2) have the form of a regular expansion in $1/N$,

$$Z_N = cS_0^{-N} \Gamma(N+b)$$
$$\times \left\{ 1 + \frac{A_1}{N} + \frac{A_2}{N^2} + \dots + \frac{A_K}{N^K} + \dots \right\}, \tag{3}$$

and calculation of these corrections provides additional information on the coefficient function $Z_N$. It was shown recently by one of the authors [2] that the series (3) diverges factorially and the typical asymptotics of coefficients $A_K$ for $K \longrightarrow \infty$ has the form

$$A_K = \tilde{c} \left( \ln \frac{S_1}{S_0} \right)^{-K} \Gamma\left( K + \frac{r'-r}{2} \right), \tag{4}$$

where $S_0$ and $S_1$ are the values of the first and second instantons in the given field theory, while $r$ and $r'$ are the corresponding numbers of zeroth modes; we assume that instantons are labelled in the order of increasing of the corresponding action. The present paper has an aim to obtain closed results for the asymptotics of $A_K$ in the $n$-component $\varphi^4$ theory with the action

$$S\{g, \varphi\} = \int d^d x \left\{ \frac{1}{2} \sum_{\alpha=1}^{n} [\nabla \varphi_\alpha(x)]^2 \right.$$
$$\left. + \frac{1}{2} m^2 \sum_{\alpha=1}^{n} \varphi_\alpha^2(x) + \frac{1}{4} g \left( \sum_{\alpha=1}^{n} \varphi_\alpha^2(x) \right)^2 \right\}, \tag{5}$$

where $d$ is the dimension of the space. Initially, we planned to calculate the constant in formula (4); however, it was found in the course of analysis that the functional form of Eq.4 should be corrected in the presence of soft modes.

Unfortunately, complete information on higher order instantons in the $\varphi^4$ theory is not available. It is known [3] that the zero-node spherically symmetric instanton derived analytically for $d = 1$, 4 and numerically for $d = 2$, 3 [4] possesses the minimal action $S_0$. Configurations with actions $2S_0$, $3S_0$, etc. corresponding to several infinitely remote elementary instantons obviously also exist; for $d = 1$, such configurations exhaust the entire instanton spectrum. Higher order spherically symmetric instantons are high-lying [1] for $d = 2$, 3 and are missing for $d = 4$. Eleonsky and co-workers [5] made an attempt at finding instantons that do not possess a spherical symmetry; two instantons with an action higher than that for the third spherically symmetric instanton were found for $d = 2$, while no nontrivial instantons were discovered for $d = 3$. Ushveridze [6] found analytically a series of asymmetric instantons for $d = 4$, which begins with[2] $8S_0$. A numerical algorithm for determining the "main sequence" of instantons was proposed in [7]. The realization of this algorithm shows[3] that lower-lying instantons of this sequence split into elementary instantons. Thus, the most probable candidate to the role of the second instanton in the $\varphi^4$ theory is a combination of two elementary instantons; the present study is mainly based on this assumption. However, since the existence of a asymmetric instanton with an action smaller than $2S_0$ cannot be ruled out, formal results corresponding to this case are given in Section 8.

---

[1] The second spherically symmetric instanton has an action $6.6S_0$ for $d = 2$ and $6.3S_0$ for $d = 3$.

It will be shown below that result (4) is valid when the "equipartition law" is applicable, i.e., when all fluctuation modes can be clearly separated into the zero and oscillatory modes (Section 2). For two-instanton configurations, a soft mode corresponding to a change in the distance between elementary instantons and reducible to oscillations at a nonanalytic minimum inevitably exists. As a result, the right-hand side of formula (4) acquires logarithmic corrections for $d = 1$, 2, 3 and even power corrections for $d = 4$.

Saddle-point calculations for two-instanton configurations were considered in [8–12] in connection with the Lipatov asymptotics for problems with degenerate vacuum (such as quantum chromodynamics). In this case, the main difficulty lies in the emergence of poorly defined integrals, which were interpreted in [8–12] at a level of heuristic recipes, and not sufficient consistency of such interpretation was admitted by the authors themselves. In the calculation of the asymptotic form of $A_K$, this problem acquires a new aspect and requires a thorough analysis. For this reason, we begin with the discussion of the Bogomolny–Parisi dispersion relation [13, 14], which is a source of poorly defined expressions (Sec.3). On the basis of this relation, the general correspondence between the corrections to the asymptotics and higher instantons is established, in confirmation of heuristical considerations of [2], (Sec.4). Then the rule for combination of instantons is derived (Sec. 5) and the general computational algorithm in the presence of soft modes is formulated (Sec. 6) and then applied to the $\varphi^4$ theory (Sec.7). The results for an asymmetric second instanton are presented in Section 8.

[2] The value $27/16S_0$ indicated in [2] is erroneous.

[3] E.P. Podolyak, private communication.

We will study functional integrals of the form

$$Z_M(g) = \int D\varphi \, \varphi_{\alpha_1}(x_1) \cdots \varphi_{\alpha_M}(x_M) \exp(-S\{g, \varphi\}), \qquad (6)$$

via which $M$-point Green functions can be expressed,

$$G_M(g) = \frac{Z_M(g)}{Z_0(g)}. \qquad (7)$$

We have derived the following expression for the asymptotic form of coefficients $A_K$ corresponding to the Green function $G_M(g)$ for $d = 1$:

$$A_K = -\frac{2^{-M/2}}{(\pi/2)\Gamma(n/2)}\left(\frac{3}{2\ln 2}\right)^{n/2}$$

$$\times \Gamma\left(K + \frac{n}{2}\right)(\ln 2)^{-K}[\ln K + C], \qquad (8)$$

$$C = C_E + \ln\left(\frac{6}{\ln 2}\right) + \frac{\psi(1/2) - \psi(n/2)}{2},$$

here, $C_E$ is the Euler constant and $\psi(x)$ is the logarithmic derivative of the gamma-function. For $d = 2$, to within the logarithmic accuracy, we have

$$A_K = -\frac{2^{-M/2}}{19.7}\frac{(0.702)^n}{\Gamma(n/2)}$$

$$\times \Gamma\left(K + \frac{n+1}{2}\right)(\ln 2)^{-K}\ln^2 K \qquad (9)$$

and, similarly, for $d = 3$ :

$$A_K = -\frac{2^{-M/2}}{2.12}\frac{(0.704)^n}{\Gamma(n/2)}$$

$$\times \Gamma\left(K + \frac{n+2}{2}\right)(\ln 2)^{-K}\ln^3 K. \qquad (10)$$

For $d = 4$, the results depend on the coordinates appearing in the Green function and are rather cumbersome (see Section 7). The expressions are simplified as we use the momentum representation and choose momenta $p_i$ corresponding to a symmetric point ($p_i \sim p$):

$$A_K = B\exp\left(\nu\ln\frac{\mu}{p}\right)$$

$$\times \Gamma\left(K + \frac{n+4}{2} + \nu\right)(\ln 2)^{-K}, \qquad (11)$$

where $\mu$ is the point of charge normalization, $\nu = (n + 8)/3$, and the values of constant $B$ are given in the table. In the scalar case ($n = 1$), the main contribution to the

$\log |A_K|$

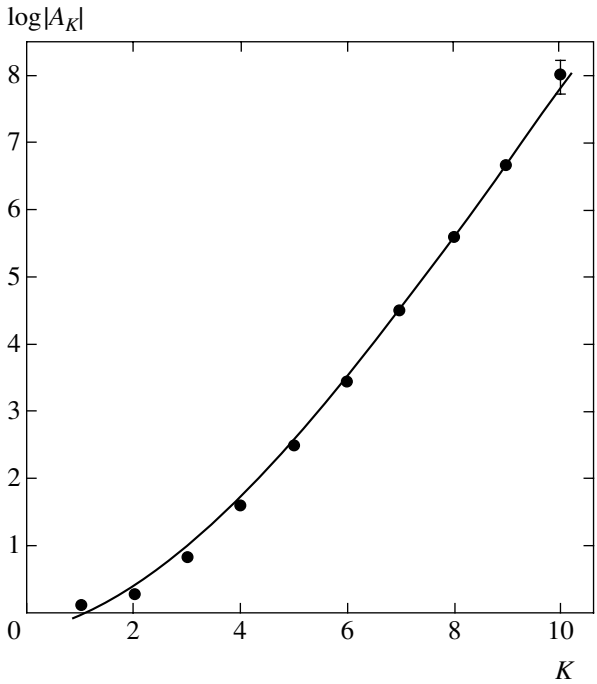

**Fig. 1.** Comparison of asymptotic formula (12) (solid curve) with coefficients $A_K$ determined numerically in [15] (circles)

asymptotics vanishes and we should expect the behavior corresponding to the next order in $1/K$:

$$A_K = \text{const} \exp\left(\nu \ln\frac{\mu}{p}\right)$$
$$\times \Gamma\left(K + \frac{n+4}{2} + \nu - 1\right)(\ln 2)^{-K}. \quad (98b)$$

The results for the logarithm of the vacuum integral $Z_0(g)$ can formally be obtained from expressions (8)–(11) by substituting $M = 0$ and introducing an additional factor $1/2$ into the right-hand sides of these expressions. In particular, for the ground-state energy of a harmonic oscillator ($d = 1$, $n = 1$), we obtain

$$A_K = -\frac{\ln K + 2.74}{3.78}\Gamma\left(K + \frac{1}{2}\right)(\ln 2)^{-K}, \quad (12)$$

which can be compared with the results obtained by

Values of parameter $B$ in formula (11)

| $n$ | $B \times 10^4$ | |
| --- | --- | --- |
| | $M = 2$ | $M = 4$ |
| 0 | −9.05 | −8.72 |
| 1 | 0 | 0 |
| 2 | 3.25 | 1.45 |
| 3 | 4.55 | 1.50 |

Bender and Wu [15] (Fig. 1); in contrast to [2], this comparison is carried out without using any fitting parameters.

If the second instanton is an asymmetric localized function (Sec.8), soft modes are absent and the structure of the result corresponds to the formula (4). For the first instanton of the $\varphi^4$ theory, zero modes include $d$ translational, $n-1$ rotational (associated with a change in the direction of the vector field $\varphi$) and (for $d = 4$) one dilatational mode corresponding to a change in the instanton radius.[4] For the second instanton, in view of its low symmetry, $d(d-1)/2$ additional modes associated with rotations in the coordinate space appear; consequently, we have

$$A_K = \tilde{c}\left[\ln\left(\frac{S_1}{S_0}\right)\right]^{-K}\Gamma\left(K + \frac{d(d-1)}{4}\right). \quad (13)$$

The latter modes have not been considered before, which makes the calculation of constant $\tilde{c}$ (Section 8) nontrivial from the methodical point of view. The method for integrating with respect to these modes is of interest for quantum electrodynamics, where even the first instanton is asymmetric [16].

## 2. STRUCTURE OF THE SADDLE-POINT CONTRIBUTION: EQUIPARTITION LAW

In the subsequent analysis, we will use the brief notation for integral (6),

$$Z(g) = \int D\varphi \, \varphi^{(1)} \ldots \varphi^{(M)} \exp(-S\{g, \varphi\}), \quad (14)$$

and will normalize it to an analogous integral with $M = 0$, $g = 0$, including factor $Z_0^{-1}(0)$ in the symbol $D\varphi$. Using the homogeneity properties of action,[5] which are typical of the $\varphi^4$ theory,

$$S\{g, \varphi\} \longrightarrow \frac{S\{\varphi\}}{g} \quad \text{for} \quad \varphi \longrightarrow \frac{\phi}{\sqrt{g}}, \quad (15)$$

and introducing the saddle-point configuration via the condition $S'\{\phi_c\} = 0$, in the vicinity of the saddle-

---

[4] For a two-instanton configuration, the number of zero modes is doubled ($r' = 2r$); by virtue of Eq. (4), this makes the contribution $r/2$ to the argument of the gamma-function, which is equal to $(n-1+d)/2$ for $d < 4$ (formulas (8)–(10)) and $(n+4)/2$ for $d = 4$ (formula (11)).

[5] Analogous homogeneity properies of action are valid in other field theories, and the subsequent analysis is also applicable to these theories after slight modifications.

point, we have

$$Z(g) = g^{-M/2} \int D\varphi \phi_c^{(1)} \ldots \phi_c^{(M)}$$

$$\times \exp\left(-\frac{S\{\phi_c\}}{g} - \frac{1}{2}(\delta\varphi, S''\{\phi_c\}\delta\varphi)\right), \qquad (16)$$

$(\delta\varphi = \varphi - \varphi_c, \varphi_c = \phi_c g^{-1/2})$; in the absence of zero modes, this leads to

$$Z(g) = \text{const}\, g^{-M/2} \exp\left(-\frac{S\{\phi_c\}}{g}\right). \qquad (17)$$

We are using the symbolic notation introduced in [2], where prime and double prime denote the first and second functional derivatives, which are treated as a vector and a linear operator, respectively, while variables $\varphi_i$ included in the symbol $D\varphi$ are assumed to be components of vector $\varphi$.

Expansion coefficients $Z_N$ are defined by the integral

$$Z_N = \int_C \frac{dg}{2\pi i g} \frac{Z(g)}{g^{N+1}}$$

$$= \int_C \frac{dg}{2\pi i g} \int D\varphi \varphi^{(1)} \ldots \varphi^{(M)} \exp\left(-\frac{S\{\phi\}}{g} - N\ln g\right), \qquad (18)$$

where contour $C$ encloses the point $g = 0$ in the positive direction. According to Lipatov [1], for large values of $N$, the integral can be evaluated by the saddle-point method. Introducing the saddle-point configuration via the conditions

$$S'\{\phi_c\} = 0, \quad g_c = \frac{S\{\phi_c\}}{N} \qquad (19)$$

and carrying out the expansion in the vicinity of this configuration, we obtain

$$Z_N = e^{-N} g_c^{-N-M/2} \int_{-\infty}^{\infty} \frac{dt}{2\pi} \int D\varphi \phi_c^{(1)} \ldots \phi_c^{(M)}$$

$$\times \exp\left(\frac{Nt^2}{2} - \frac{1}{2}(\delta\varphi, S''\{\phi_c\}\delta\varphi)\right), \qquad (20)$$

where $g = g_c + ig_c t$; in the absence of zero modes, this gives

$$Z_N = \text{const}\, S\{\phi_c\}^{-N} \Gamma\left(N + \frac{M}{2}\right).$$

In fact, the functional integral always contains zero modes; for correct integration with respect to these modes, we introduce collective variables $\lambda_i$ (such as the center of an instanton, its orientation, etc.), which are formally determined for an arbitrary configuration of $\varphi$ and are its functionals, $\lambda_i = f_i\{\varphi\}$. The latter can be treated as homogenious in $\varphi$ with zero degree of homogeneity [2]. We introduce into integral (14) the partition of unity,

$$1 = \prod_{i=1}^{r} \int d\lambda_i \delta(\lambda_i - f_i\{\varphi\})$$

$$= \prod_{i=1}^{r} \int d\lambda_i \delta(\lambda_i - f_i\{\phi_c\} - (f'\{\phi_c\}, \delta\varphi)) \qquad (21)$$

$$= \prod_{i=1}^{r} \int d\lambda_i \delta(\lambda_i - f_i\{\phi_c\} - \sqrt{g}\{f'\{\phi_c\}, \delta\varphi\}),$$

where $r$ is the number of zero modes. Using the degrees of freedom corresponding to zero modes, we choose the instanton from the condition $\lambda_i = f_i\{\phi_c\}$, after which $\phi_c$ becomes a function of $\lambda_i$ (i.e., $\phi_c \equiv \phi_\lambda$). Simple calculations lead to the following results (see [2] for details):

$$Z(g) = c_g g^{-(M+r)/2} \exp\left(-\frac{S\{\phi_c\}}{g}\right), \qquad (22)$$

$$c_g = \sqrt{\frac{\det S''\{0\}}{\det[S''\{\phi_c\}]_P}} \frac{(2\pi)^{-r/2}}{\det[f'\{\phi_c\}]_P}$$

$$\times \int \prod_{i=1}^{r} d\lambda_i \phi_\lambda^{(1)} \ldots \phi_\lambda^{(M)}, \qquad (23)$$

$$Z_N = c S\{\phi_c\}^{-N} \Gamma\left(N + \frac{M+r}{2}\right), \qquad (24)$$

$$c = \frac{S\{\phi_c\}^{-(M+r)/2}}{(2\pi)^{1+r/2}} \sqrt{-\frac{\det S''\{0\}}{\det[S''\{\phi_c\}]_{P'}}}$$

$$\times \frac{1}{\det[f'\{\phi_c\}]_P} \int \prod_{i=1}^{r} d\lambda_i \phi_\lambda^{(1)} \ldots \phi_\lambda^{(M)}, \qquad (25)$$

where $f'\{\phi_c\}$ is the operator whose matrix consists of columns $f_i'\{\phi_c\}$, while subscripts $P$ and $P'$ mark the projection onto the subspace of zero modes and the supplementary subspace, respectively.[6] Expression (24) reproduces the functional form of the Lipatov asymptotics (2) given above.

In accordance with relation (24), each degree of freedom corresponding to zero modes gives $1/2$ in the argument of the gamma function. This resembles the classicalcal equipartition law, and in fact (under the

---

close consideration) is in exact correspondence with it. As a matter of fact, the classical partition function $Z$ is determined by a configurational integral of $\exp(-H/T)$; as the number $r_{osc}$ of the oscillatory degrees of freedom increases by unity, the substitution $Z \longrightarrow ZT^{1/2}$ takes place, which makes an additional contribution of $1/2$ to heat capacity [17]. Integral (14) we are interested in is determined by the exponential $\exp(-S\{\phi\}/g)$, where the coupling constant $g$ plays the role of temperature. An increase in the number $r$ of zero modes by unity corresponds to a decrease in $r_{osc}$ by unity and leads to the substitution $Z \longrightarrow Zg^{-1/2}$ (see relation (22)). In the calculation of the Lipatov asymptotics, factor $g^{-1/2}$ is estimated at the saddle point $g_c \sim 1/N$ (see relation (19)), which leads to the substitution $Z \longrightarrow Z_N N^{1/2}$ and to the addition of $1/2$ to the argument of the gamma function.

The equipartition law can be violated in the presence of soft modes associated with approximate symmetries of the system. In this case, some degrees of freedom in the first approximation appear as zero modes; however, in a more accurate analysis, these degrees of freedom correspond to motion in a potential relief that may be irreducible to a quadratic minimum. Examples of soft modes are dilatations in the massive four-dimensional or $(4 - \epsilon)$-dimensional $\varphi^4$ theory [18, 19] and the change in the distance between elementary instantons in a two-instanton configuration (see below).

## 3. DISPERSION RELATION AND ROTATION RULE

In view of the factorial growth of coefficients $Z_N$, series (1) has zero radius of convergence. This is due to the fact that point $g = 0$ is a branching point; to single out the regular branch of $Z(g)$, we should make a cut in the complex $g$ plane from 0 to $\infty$ along the ray

$$g = |g| \operatorname{sgn} S_0, \qquad (26)$$

where the Borel sum of series (1) is poorly defined.[7] Using the Cauchy formula and writing it in the form of a dispersion relation, we obtain

$$Z(g) = \frac{1}{2\pi i} \oint_C \frac{Z(\tilde{g})}{\tilde{g} - g} d\tilde{g} = \frac{1}{2\pi i} \int_0^{\infty \operatorname{sgn} S_0} \frac{\Delta Z(\tilde{g})}{\tilde{g} - g} d\tilde{g}. \qquad (27)$$

$$\Delta Z(g) = Z(g + i0 \operatorname{sgn} S_0) - Z(g - i0 \operatorname{sgn} S_0), \qquad (28)$$

where contour $C$ embraces the point $\tilde{g} = g$ and then is deformed so as to enclose the cut. Expanding func-

---

[7] In the $\varphi^4$ theory, action $S_0$ is negative and the cut is made along the negative semiaxis.

tion (27) into a series in $g$, we obtain the following expression for the expansion coefficients:

$$Z_N = \frac{1}{2\pi i} \int_0^{\infty \operatorname{sgn} S_0} \frac{\Delta Z(g)}{g^{N+1}} dg. \qquad (29)$$

The asymptotic form of $Z_N$ for $N \longrightarrow \infty$ and the discontinuity at the cut for $g \longrightarrow 0$ are connected through the relation

$$\Delta Z(g) = 2\pi i c \left(\frac{S_0}{g}\right)^b \exp\left(-\frac{S_0}{g}\right)$$

$$\longleftarrow Z_N = c S_0^{-N} \Gamma(N + b), \qquad (30)$$

which can easily be derived from formula (29) or by calculating the discontinuity at the cut of the Borel sum of series (1). The next step is to identify $\Delta Z(g)$ with the result of the saddle-point evaluation of integral $Z(g)$ in the vicinity of the same configuration $\varphi_c$ as in the Lipatov method,

$$\Delta Z(g) = [Z(g)]_{\text{saddle-point } \phi_c}. \qquad (31)$$

Relations (30) and (31) were proposed by Bogomolny [13] and Parisi [14] and form the basis of the approach to calculating higher orders, which is alternative to the Lipatov method. These relations enable us to easily find the asymptotic form of $Z_N$ if the result of the saddle-point calculation of $Z(g)$ is already known. Relation (31) can be substantiated for a conventional integral using the elegant analysis proposed by Langer [12, 20]; however, this relation has never been proved in the general form; besides, it is poorly defined and requires an appropriate interpretation.

To substantiate relation (31), we must formulate the rule of permutation of integrations with respect to $g$ and $\varphi$; we introduce this rule using as an example the conventional integral:

$$Z(g) = \int_0^\infty d\varphi \exp(-\varphi^2 - g\varphi^4), \qquad (32)$$

$$Z_N = \int_C \frac{dg}{2\pi i g} \int_0^\infty d\varphi \exp(-\varphi^2 - g\varphi^4 - N \ln g). \qquad (33)$$

Diverging series (1) is obtained as a result of a regular expansion of the exponent in relation (32) in $g$ followed by interchanging summation and integration; as for the direct expansion of $Z(g)$ into a series, it is not quite correct since it corresponds to the Taylor expansion at a clearly singular point. For this reason, it is appropriate in expression (33) to integrate first with respect to $g$ (determining the coefficient of the expansion of $\exp(-g\varphi^4)$) and then with respect to $\varphi$; the con-tour C can be chosen circular so that it passes through the saddle point $g_c = -1/N$ in the vertical direction (Fig. 2a). Interchanging the sequence of integrations is not quite trivial in view of the presence of the cut in $Z(g)$ along the ray (2). Let us restrict integration over the domain of large $\varphi$ values using a truncating multiplier to shift the cut by a distance $\Delta$ from zero and deform the contour as shown in Fig. 2b. If we expand the contour and remove the truncation, the cut will be circumvented in the negative direction; it can easily be seen that this corresponds to formula (29).[8]

Depending on the sequence of integration, the saddle point $g_c$ is passed by either in the vertical (see Fig. 2a) or in the horizontal direction (Fig. 2b). It is well known that this point should be passed in the steepest descend direction. Whether this direction is vertical or horizontal?

In integral (33), the saddle point takes place at $\varphi_c = \sqrt{2N}$, $g_c = -1/N$ and the quadratic form appearing in the expansion in the vicinity of this point ($\varphi = \varphi_c + \delta\varphi$, $g = g_c + itg_c\sqrt{2/N}$) can be written as

$$-(\delta\varphi)^2 - (t - i\sqrt{2}\delta\varphi)^2 \ \text{ or } \ t^2 + (\delta\varphi + i\sqrt{2}t)^2. \quad (34)$$

If we integrate first with respect to $g$, the shift in variable $t$ leads to the well-defined Gaussian integral

$$\int\limits_{-\infty}^{\infty} d\varphi \int\limits_{-\infty}^{\infty} dt \exp[-(\delta\varphi)^2 - (t - i\sqrt{2}\delta\varphi)^2]$$

$$= \int\limits_{-\infty}^{\infty} d\varphi \int\limits_{-\infty}^{\infty} dt \exp[-(\delta\varphi)^2 - t^2]. \quad (35)$$

If we integrate first with respect to $\varphi$, we obtain a "bad" integral

$$\int\limits_{-\infty}^{\infty} dt \int\limits_{-\infty}^{\infty} d\varphi \exp[(\delta\varphi + i\sqrt{2}t)^2 + t^2]$$

$$= \int\limits_{-\infty}^{\infty} dt \int\limits_{-\infty}^{\infty} d\varphi \exp[(\delta\varphi)^2 + t^2]. \quad (36)$$

Let us turn the contours of integration with respect to $t$ and $\varphi$ in Eq. (36) through the same angle in the opposite directions ($t \longrightarrow te^{-i\alpha}$, $\delta\varphi \longrightarrow \delta\varphi e^{i\alpha}$); this does not change the determinant of the quadratic form, which defines the Gaussian integral. For $\alpha = \pi/2$, integral (36) is transformed into (35) and a transition from Fig. 2c to Fig. 2d takes place: the vertical integration with respect to $g$ becomes horizontal, as is required by formula (29), while the discontinuity at the cut $\Delta Z(g)$ is obtained from the initial integral $Z(g)$ by rotating the contour relative to the saddle point $\varphi_c$ through an angle $\pi/2$ in the positive direction.

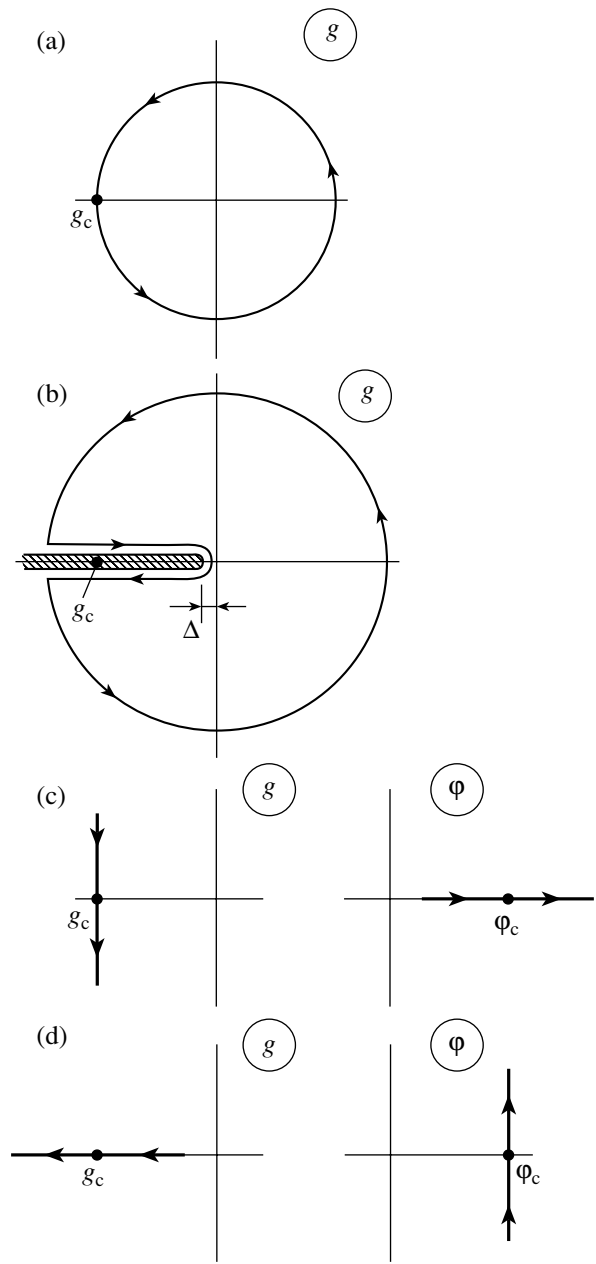

**Fig. 2.** (a) Contour of integration over $g$ in expression (33) can be chosen in the form of a circle if we first integrate with respect to $g$; (b) contour of integration over $g$ after changing the sequence of integrations; (c, d) contours of integration over $g$ and $\varphi$ are rotated in the opposite directions in the course of changing the sequence of integrations.

---

[8] At first glance, another line of reasoning is possible. Let us deform contour $C$ so that it embraces the negative semiaxis, so the cut will appear within it; in this case we obtain circumvention of the cut in the positive direction. It is suggested in this reasoning that the cut "grows" from zero to infinity. This corresponds to the Taylor expansion at the singular point and, hence, is incorrect.

This "rotation rule" can easily be extended to the general case. The main contribution to integral (18) comes from the configuration of $\phi_c$ corresponding to the maximum of the integrand. Consequently, the quadratic form in the exponential in Eq. (20) should lead to the sum of squares, which is possible for

$$-N \det S''\{\phi_c\} > 0. \tag{37}$$

However, integral (16) for $Z(g)$ in this case is poorly defined: it contains a "bad" Gaussian integration, which should be understood in the sense

$$\int\limits_{-\infty}^{\infty} dx\, e^{x^2} \longrightarrow \int\limits_{-\infty}^{\infty} d(ix)\ e^{-x^2}. \tag{38}$$

After that integral (16) defines the discontinuity at the cut $\Delta Z(g)$. "Bad" integration can be regarded as unique without loss of generality, because it is possible to reverse simultaneously the signs before the squares of two variables. Thus, the square root of the determinant in Eq. (23) should obviously be treated as

$$(\det[S''\{\phi_c\}]_P)^{-1/2} \longrightarrow i \left|\det[S''\{\phi_c\}]_P\right|^{-1/2}, \tag{39}$$

after which the results (22) and (24) satisfy relation (30).

The rotation rule solves the problem of interpretation of the discontinuity $\Delta Z(g)$ at the cut in the framework of the Gaussian approximation, while additional analysis is required in the presence of soft modes (Section 6).

## 4. RELATION BETWEEN CORRECTIONS TO ASYMPTOTICS AND HIGHER INSTANTONS

Separating the Lipatov asymptotics from coefficients $Z_N$, we obtain, in accordance with expression (3),

$$Z_N = c S_0^{-N} \Gamma\left(N + \frac{M+r}{2}\right) F\left(\frac{1}{N}\right), \tag{40}$$

and coefficients $A_K$ can be expressed in terms of function $F(\epsilon)$:

$$A_K = \int\limits_C \frac{d\epsilon}{2\pi i} \frac{F(\epsilon)}{\epsilon^{K+1}}. \tag{41}$$

Substituting expression (18) into (40), setting $\epsilon = 1/N$, and carrying out the substitution $g \longrightarrow \epsilon S_0 g$, we obtain

the exact expression for $A_K$:

$$A_K = \frac{1}{\sqrt{2\pi}c} \int \frac{d\epsilon}{2\pi i \epsilon} \int \frac{dg}{2\pi i g} \int D\phi\, \phi^{(1)} \ldots \phi^{(M)} \epsilon^{(M+r-1)/2}$$
$$\times \exp\left\{\frac{1}{\epsilon}\left[1 - \ln g - \frac{S\{\phi\}}{S_0 g}\right] - K \ln \epsilon\right\}_{\phi = \phi_c \sqrt{\epsilon g S_0}}, \tag{42}$$

which can be estimated by the saddle-point method for large values of $K$. The saddle-point configuration is defined by the conditions

$$S'\{\psi_c\} = 0, \quad g_c = \frac{S'\{\psi_c\}}{S_0}, \quad \epsilon_c = \frac{\ln g_c}{K}, \tag{43}$$

and expansion in the vicinity of this configuration gives the exponential in the form

$$\exp\left\{-\frac{1}{2}\left[(\delta\varphi, S''\{\psi_c\}\delta\varphi) - \frac{t^2}{\epsilon_c} + K\tau^2\right]\right\}, \tag{44}$$

where $g - gc = ig_c t$ and $\epsilon - \epsilon_c = i\epsilon_c \tau$. For $\psi_c$, we cannot use the configurations $\psi_c = 0$ and $\psi_c = \phi_c$ since $g_c = 0$ or $\epsilon_c = 0$ in this case; this corresponds to a singularity rather than a saddle point. The saddle point corresponds to the maximum of the integrand under the condition

$$\frac{K}{\epsilon_c} \det S''\{\psi_c\} > 0, \tag{45}$$

and we should take the first of the higher instantons for which $\det S''\{\psi_c\} > 0$ for $\psi_c$. This condition is usually satisfied already for the second instanton, and this will be presumed in the subsequent analysis. Evaluation of integral (42) gives formula (4), in which

$$\tilde{c} = \frac{(\ln S_1/S_0)^{(r-r')/2} S_1^{-(M+r')/2}}{c(2\pi)^{2+r'/2} \det[f'\{\psi_c\}]_P}$$
$$\times \sqrt{\frac{\det S''\{0\}}{\det[S''\{\psi_c\}]_P}} \prod_{i=1}^{r'} d\lambda_i \psi_\lambda^{(1)} \ldots \psi_\lambda^{(M)} \tag{46}$$

and $S_1 = S\{\psi_c\}$ (see [2] for details).

The dispersion relation (see Section 3) enables us to establish the correspondence between the corrections to the asymptotics and higher instantons in the general form without resorting to specific features of the $\varphi^4$ theory. Let coefficients $A_K$ increase for large values of $K$ according to the factorial law

$$A_K = \tilde{c}\tilde{a}^K \Gamma(K + \tilde{b}) \tag{47}$$

with $\tilde{a} > 0$; in this case, the Borel sum of the series in Eq. (3) is poorly defined for $N > 0$ and the coefficient function $Z_N$ has a cut with a discontinuity at it defined by rule (30):

$$\Delta Z_N = c S_0^{-N} \Gamma(N+b) \times 2\pi i \tilde{c} \left(\frac{N}{\tilde{a}}\right)^{\tilde{b}} \exp\left(-\frac{N}{\tilde{a}}\right). \quad (48)$$

On the other hand, this discontinuity can be defined in analogy with (31) as the contribution of the saddle-point configuration $\psi_c$ to the Lipatov integral (18). Assuming that the functional form of this contribution is analogous to (2),

$$\Delta Z_N = i c_1 S_1^{-N} \Gamma(N+b_1), \quad (49)$$

and identifying expression (48) with (49), we obtain

$$\tilde{a} = \left(\ln\frac{S_1}{S_0}\right)^{-1}, \quad \tilde{b} = b_1 - b,$$
$$\tilde{c} = \frac{c_1}{2\pi c}\left(\ln\frac{S_1}{S_0}\right)^{b-b_1}. \quad (50)$$

As a result, the determination of the parameters of the asymptotic form of $A_K$ can be reduced to the well-known procedure: it is sufficient to evaluate the saddle-point contributions (2) and (49) to the Lipatov integral (18) from two configurations, $\phi_c$ and $\psi_c$. Result (46) readily follows from (50) if we take into account the fact that expression (49) can be derived from (25) via the substitution $\phi_c \longrightarrow \psi_c$, $r \longrightarrow r'$, and $(-\det[S''\{\psi_c\}]_P)^{-1/2}$ is treated as $i|\det[S''\{\psi_c\}]_P|^{-1/2}$ in accordance with the rotation rule.

Formulas (50) solve the problem of evaluation of the asymptotic form of $A_K$ under the condition of applicability of the equipartition law; for this purpose, all fluctuational modes in the vicinity of classical configurations $\phi_c$ and $\psi_c$ should be distinctly separable into zero and oscillatory modes. In this case, $b_1 - b = (r'-r)/2$ irrespective of the specific features of the $\phi^4$ theory: a contribution of the $M/2$ type in the argument of the gamma-function, which stems from the preexponential factor in formula (14), may have different values in other field theories, but it is the same for the first and second instantons. In the presence of soft modes, the situation is more complicated and will be considered in the following sections.[9]

## 5. INSTANTON COMBINATION RULE

Let us find out how to construct the contribution from a two-instanton configuration, knowing the contribution from one instanton to the functional integral (14). In accordance with Section 2, the contribution to $Z_M(g)$ from the saddle-point configuration $\phi_c$ has the structure

$$Z_M^{(1)}(g) = c_0 g^{-(M+r)/2} \exp\left(-\frac{S_0}{g}\right) \int \prod_{i=1}^{r} d\lambda_i \phi_\lambda^{(1)} \ldots \phi_\lambda^{(M)}, \quad (51)$$

where $c_0 = c_0\{\phi_c\}$ and $S_0 = S\{\phi_c\}$ are functionals of $\phi_c$. The contribution from the two-instanton configuration is defined by an analogous expression, in which $\phi_c$ is replaced by $\phi_\lambda + \phi_{\lambda'}$. If we introduce the instanton interaction $S_{int}$ by the relation

$$S\{\phi_\lambda + \phi_{\lambda'}\} = S\{\phi_\lambda\} + S\{\phi_{\lambda'}\} + S_{int}\{\phi_\lambda, \phi_{\lambda'}\} \quad (52)$$

and take into account the doubling of the number of collective variables, we obtain the sum of terms of the form

$$Z_{LL'} = c_1 g^{-M/2-r} \exp\left(-\frac{2S_0}{g}\right) \int \prod_{i=1}^{r} d\lambda_i d\lambda'_i \phi_\lambda^{(1)} \ldots \phi_\lambda^{(L)}$$
$$\times \phi_{\lambda'}^{(1)} \ldots \phi_{\lambda'}^{(L')} \exp\left(-\frac{S_{int}\{\phi_\lambda, \phi_{\lambda'}\}}{g}\right) \quad (53)$$

where $L + L' = M$. For small values of $g$, the exponential restricts the interaction between instantons by the condition $S_{int}\{\phi_\lambda, \phi_{\lambda'}\} \lesssim g$; consequently, we can disregard this interaction in the preexponential factor. It can naturally be expected that $c_1$ must be equal to $c_0^2$ for expression (53) at $S_{int} \equiv 0$ to be just the product of the right-hand sides of Eqs. (51) with $M = L$ and $M = L'$. Small values of $S_{int}$ correspond to remote instantons,[10] which enables us to disregard cross terms containing simultaneously $\phi_\lambda$ and $\phi_{\lambda'}$. In this case, the sum over $L, L'$ will contain only two terms with $L = M$, $L' = 0$ and $L = 0$, $L' = M$, which are obviously identical. The emerging factor 2 is cancelled out with the combinatorial multiplier $1/2!$, which should be introduced in view of the fact that configurations differing in the permutation of instantons are taken into account twice. As a result, the two-instanton contribution assumes the form

$$Z_M^{(2)}(g) = c_0^2 g^{-M/2-r} \exp\left(-\frac{2S_0}{g}\right)$$
$$\times \int \prod_{i=1}^{r} d\lambda_i d\lambda'_i \phi_\lambda^{(1)} \ldots \phi_\lambda^{(M)} \exp\left(-\frac{S_{int}\{\lambda, \lambda'\}}{g}\right), \quad (54)$$

where the fact that $S_{int}\{\phi_\lambda, \phi_{\lambda'}\}$ for a fixed formed of instantons depends only on $\lambda$ and $\lambda'$ is taken into account. For $M = 0$, configurations with $L = M$, $L' = 0$ and $L = 0$, $L' = M$ coincide and the result is defined by formula (54) with an additional factor $1/2$ on the right-hand side; this remark applies to all subsequent expressions as well.

---

[9] Relations of type (50) are valid in this case as well (with allowance for possible change in the functional form of expressions (47) and (49)); however, these relations are practically useless since the arising integrals are poorly defined.

[10] See modifications for $d = 4$ in Sec. 7.

Formula (54) enables us to write the expression for the two-instanton contribution on the base of the known result (51) for the one-instanton contribution; we only require additional information about the interaction of instantons at large distances. The rule for combination of instantons appears as quite natural, but some subtle aspects missing in heuristic derivation should be discussed.

**Introducing of a constraint.** In fact the replacement of $\phi_c$ by $\phi_\lambda + \phi_{\lambda'}$ is not quite correct since the linear combination of instantons is not the exact solution of the equation $S'\{\phi\} = 0$. For this reason, the expansion in the vicinity of this configuration acquires the terms which are linear in $\delta\phi$ and require accurate elimination.[11]

Let us introduce the collective variable $z$ characterizing the distance $R$ between the instantons and formally defined for an arbitrary instanton configuration, $z = f\{\phi\}$. The idea lies in finding the extremum of action under an additional condition (constraint) $f\{\phi\} = \text{const}$ (i.e., for a fixed distance between instantons) and in subsequent integrating with respect to this distance. In this case, the instanton is defined by the equation

$$S'\{\phi_c\} - \mu f'\{\phi_c\} = 0 \qquad (55)$$

($\mu$ is the Lagrange multiplier) and the integration with respect to $z$ is carried out by introducing the partition of unity,

$$
\begin{aligned}
1 &= \int dz \, \delta(z - f\{\phi\}) \\
&= \int dz \, \delta(z - f\{\phi_c\} - (f'\{\phi_c\}, \delta\phi))
\end{aligned}
\qquad (56)
$$

into the functional integral. Choosing the instanton from the condition $z = f\{\phi_c\}$, we obtain

$$Z(g) = \int D\phi \, \phi^{(1)} \ldots \phi^{(M)}$$

$$\times \exp\left\{-\frac{S\{\phi_c\} + (S'\{\phi_c\}, \delta\phi) + (\delta\phi, S''\{\phi_c\}\delta\phi)}{g}\right\}(57)$$

$$\times \int dz \, \delta(-(f'\{\phi_c\}, \delta\phi))$$

and the terms linear in $\delta\phi$ in the exponential are eliminated by the $\delta$-function in view of condition (55). For $f\{\phi\}$, it is convenient to take the quantity $S_{\text{int}}\{\phi_\lambda, \phi_{\lambda'}\}$ since Eq. (55) has a combination of instantons $\phi_\lambda + \phi_{\lambda'}$ as the exact solution for $\mu = 1$ (cf. formula (52)).

We can disregard the instanton interaction in the preexponential factor; in this case, zero modes are treated in the same manner as if they were independent. The only subtle point is that, instead of zero modes

$$\frac{\partial \phi_\lambda}{\partial x_1} \quad \text{and} \quad \frac{\partial \phi_{\lambda'}}{\partial x_1}, \qquad (58)$$

corresponding to translations of instantons along the straight line passing through their centers (which is chosen as the $x_1$ axis), we should take their linear combinations

$$\frac{\partial \phi_\lambda}{\partial x_1} + \frac{\partial \phi_{\lambda'}}{\partial x_1} \quad \text{and} \quad \frac{\partial \phi_\lambda}{\partial x_1} - \frac{\partial \phi_{\lambda'}}{\partial x_1}, \qquad (59)$$

corresponding to translation of a pair as a whole and to a change in the spacing between the pair components. In this case, the $\delta$-function corresponding to the second mode in (59) is not introduced into product (21) since its role is played by those in (56). This modification is of no importance since $\det[f'\{f_c\}]_P$ in expressions (23) and (25) is in fact determined by the Gram matrix constructed on zero modes (Section 8) and is independent of the choice of functionals $f_i\{\phi\}$. Consequently, the final result (44) correspond to the formal substitution $\phi_c \longrightarrow \phi_\lambda + \phi_{\lambda'}$.

**Factorization of determinants.** While deriving formula (54), we assumed that $c_1 = c_0^2$, or

$$c_0\{\phi_\lambda + \phi_{\lambda'}\} = c_0\{\phi_\lambda\} c_0\{\phi_{\lambda'}\}. \qquad (60)$$

To clarify this relation, we note that operator $S''\{\phi_c\}$ in the scalar case has the form of the Schrödinger operator (Section 8)

$$S''\{\phi_c\} = -\Delta + m^2 - 3\phi_c^2(x), \qquad (61)$$

and, instead of a single potential well, two potential wells separated by a large distance appear upon the substitution $\phi_c \longrightarrow \phi_\lambda + \phi_{\lambda'}$. Consequently, the eigenvalues of operator $S''\{\phi_\lambda + \phi_{\lambda'}\}$ are doubly degenerate eigenvalues of operator $S''\{\phi_c\}$ and

$$\det S''\{\phi_\lambda + \phi_{\lambda'}\} = [\det S''\{\phi_c\}]^2 \qquad (62)$$

under the condition that the product of eigenvalues converges. Such convergence does not take place for operator $S''\{\phi_c\}$ and the normalization of integral (14) to quantity $Z_0(0)$ is essential. As a result of this normalization, the combination

$$\frac{\det S''\{\phi_c\}}{\det S''\{0\}} = \det \frac{S''\{\phi_c\}}{S''\{0\}}, \qquad (63)$$

appears, where operator $S''\{\phi_c\}/S''\{0\}$ has a discrete spectrum [21] and the convergence of the products of eigenvalues is ensured by a simple renormalization (see [4] for details). The multipliers singled out from relation (63) as a result of renormalization and during the elimination of zero modes as a result of the substitution $\phi_c \longrightarrow \phi_\lambda + \phi_{\lambda'}$ are squared in view of the equivalence of two instantons. Factorization of $\det[f'\{\phi_\lambda + \phi_{\lambda'}\}]_P$ similar to (62) is due to the fact that zero modes are considered in the approximation of noninteracting instantons.

---

[11] The instanton interaction depends to a considerable extent on the specific choice of the two-instanton configuration; with an inappropriate choice, the results can easily be erroneous (cf. [8–12]).

It should be noted that factorization (60) is not valid for the combination of topological instantons connecting degenerate but nonequivalent vacuums [8, 12]. In this case, the effective potential appearing in the Schrödinger operator of type (61) does not have a form of two isolated potential wells: a potential barrier emerging between the wells makes the interaction between the instanton being of a long-range type.

**Subtraction of the ideal gas contribution.**
As we transfer to the Green function, result (54) assumes the form

$$
G_M^{(2)}(g) = c_0^2 g^{-M/2 - r} \exp\left(-\frac{2S_0}{g}\right)
$$
$$
\times \int \prod_{i=1}^r d\lambda_i d\lambda_i' \phi_\lambda^{(1)} \dots \phi_\lambda^{(M)} \left[ \exp\left(-\frac{S_{\text{int}}\{\lambda, \lambda'\}}{g}\right) - 1 \right], \tag{64}
$$

i.e., the contribution of noninteracting instantons is subtracted in the same way as in the case of the virial expansion [17]. Indeed, we can write the Green function in the form

$$
G_M(g) = \frac{Z_M(g)}{Z_0(g)}
$$
$$
= \frac{Z_M^{(0)}(g) + Z_M^{(1)}(g) + Z_M^{(2)}(g) + \dots}{1 + Z_0^{(1)}(g) + Z_0^{(2)}(g) + \dots}, \tag{65}
$$

where the superscripts "0," "1," "2," ... mark the contributions from saddle-point configurations[12] $\phi = 0$, $\phi = \phi_c$, $\phi = \phi_\lambda + \phi_{\lambda'}$, ... and having the order 1, $\exp(-S_0/g)$, $\exp(-2S_0/g)$, ... . For convenience, the normalization of functional integrals in this expression is carried out not to $Z_0(0)$, but to $Z_0^{(0)}(g)$, which is immaterial in the main order in $g$. The contribution to $G_M(g)$ of the order $\exp(-2S_0/g)$ has the form

$$
Z_M^{(2)}(g) - Z_M^{(1)}(g) Z_0^{(1)}(g) - Z_M^{(0)}(g) Z_0^{(2)}(g), \tag{66}
$$

where the last term is small for $M > 0$ due to additional factor $g^{M/2}$. It can easily be seen that the second

---

[12] Such an "expansion in instantons" in the case of an ordinary integral is obtained by deforming the integration contour in such a way that it can be presented as the sum of integrals over contours $C_0$, $C_1$, $C_2$, ..., where each contour $C_i$ passes through the saddle point $z_i$ in the steepest descent direction and its ends go to infinity.

---

term corresponds to expression (54) with $S_{\text{int}} \equiv 0$, which gives $-1$ in formula (64). Similarly, after taking logarithm of the vacuum integral,

$$
\ln Z_0^{(0)}(g) = \ln[1 + Z_0^{(1)}(g) + Z_0^{(2)}(g) + \dots]
$$
$$
= Z_0^{(1)}(g) + Z_0^{(2)}(g) - \frac{1}{2}[Z_0^{(1)}(g)]^2 + \dots, \tag{67}
$$

the last term leads to the subtraction of the contribution from noninteracting instantons in the expression for $Z_0^{(2)}(g)$; in this case, the presence of the additional factor $1/2$ is significant in expressions with $M = 0$.

## 6. HIGH-ORDER CORRECTIONS TO ASYMPTOTICS IN THE PRESENCE OF SOFT MODES

Let us analyze expression (64) in the case of a power interaction between instantons separated by large distances:

$$
S_{\text{int}}\{\lambda, \lambda'\} = \frac{a(\lambda, \lambda')}{R^\alpha}. \tag{68}
$$

Introducing into expression (64) the partition of unity,

$$
1 = \int_0^\infty dz\, \delta\left(z - \frac{S_{\text{int}}\{\lambda, \lambda'\}}{2S_0}\right)
$$
$$
+ \int_0^\infty dz\, \delta\left(z + \frac{S_{\text{int}}\{\lambda, \lambda'\}}{2S_0}\right), \tag{69}
$$

and considering that integration with respect to collective variables includes the integral $\int R^{d-1} dR$, we obtain

$$
\int \int \prod_{i=1}^r \delta\lambda_i \delta\lambda_i' \phi_\lambda^{(1)} \dots \phi_\lambda^{(M)} \delta\left(z \pm \frac{S_{\text{int}}\{\lambda, \lambda'\}}{2S_0}\right) = \frac{A^\pm}{z^{1+\nu}}, \tag{70}
$$

where $\nu = d/\alpha$. As a result, expression (64) takes the functional form

$$
G^{(2)}(g) = B\left(\frac{2S_0}{g}\right)^{(M + 2r)/2} \exp\left(-\frac{2S_0}{g}\right) F\left(\frac{2S_0}{g}\right), \tag{71}
$$

where

$$
F(x) = A^+ I^+(x) + A^- I^-(x), \tag{72}
$$

$$
I^\pm(x) = \int_0^\infty \frac{dz}{z^{1+\nu}} (e^{\pm xz} - 1). \tag{73}
$$

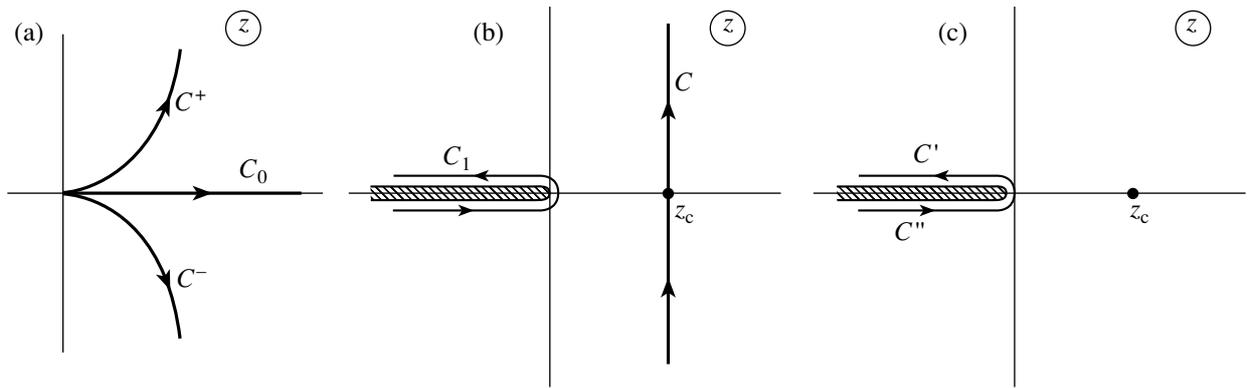

**Fig. 3.** (a) Integration contour $C_0$ in integral $I^+(x)$ is bent upwards or downwards upon a displacement of $g$ to the complex plane;

(b) integration contour $C$ defines the discontinuity at the cut of integral $I^+(x)$; to evaluate the integral, the contour is deformed to $C_1$;

(c) in the calculation of the asymptotic form of $A_K$, half-sum of the integrals over $C'$ and $C''$ appears. [direction of the latter is shown wrongly]

We will henceforth assume that $\nu \geq 0$ since negative values of $\nu$ correspond to nonphysical increase of the interaction with th e distance. For $\nu \geq 1$, subtraction of unity does not ensure the convergence of the integral in formula (73). Such values of $\nu$ correspond to a slowly decreasing interaction, for which the virial expansion is not applicable. The interaction for large values of $R$ should be modified in the spirit of Debye screening, which leads to truncation of the integral for small values of $z$. The type of truncation is immaterial since integral $\Gamma_{(x)}$ can be evaluated by differentiating with respect to parameter $x$; for $F(x)$, we formally get

$$F(x) = \Gamma(-\nu)[A^- x^\nu + A^-(-x^\nu) + O(x^{[\nu]})], \quad (74)$$

where $[\ldots]$ is the integral part of the number. The values of $g$ near the cut(where $x = 2S_0/g > 0$) are essential and expression (74) is ill-defined in view of integral $I^+(x)$ being poorly defined . The interpretation of this expression depends on the formulation of the problem; two versions of such interpretation will be considered below.

**Contribution to the Lipatov asymptotics.** Analysis of the contribution from a two-instanton configuration to the asymptotic form of coefficients $Z_N$ is important for problems with degenerate vacuum, in which a solitary instanton is topological and the Lipatov asymptotics is determined by the instanton–antiinstanton pair [8–12]. In view of formula (71), the discontinuity at the cut of function $G(g)$ is defined by the formula

$$\Delta G(g) = \pm B\left(\frac{2S_0}{g}\right)^{(M+2r)/2} \exp\left(-\frac{2S_0}{g}\right)$$
$$\times \left[F\left(\frac{2S_0}{g} - i0\right) - F\left(\frac{2S_0}{g} + i0\right)\right], \quad (75)$$

where we represent $G(g)$ as an "expansion" in instantons $G^{(0)}(g) + G^{(1)}(g) + G^{(2)}(g) + \ldots$ (see Section 4) and assume that $G^{(0)}(g)$ and $G^{(1)}(g)$ make zero contributions to the discontinuity at the cut. The indefinite sign in formula (75) is due to the fact that the expansion in instantons requires a preliminary deformation of the integration contour so that it passes through all saddle points (see footnote 12): in the general case, such a deformation is ambiguous and leads to the indefinite sign of $G^{(2)}(g)$ in view of the possibility to pass through a saddle point in two opposite directions. It can easily be seen that the well-defined integral $I^-(x)$ makes zero contribution to discontinuity (75), while in expression for $I^+(x)$, we can omit the term with $-1$. As $g$ is shifted to the lower or upper half-plane, the contour of integration with respect to $z$ in integral $I^+(x)$ should be bent upwards or downwards (Fig. 3a). The discontinuity at the cut is determined by the difference in the integrals over contours $C^+$ and $C^-$ and can be reduced to the vertical contour $C$ (Fig. 3b).

To fix the sign in expression (75), we should establish its relation with the rotation rule (see Section 3). We replace $R^{-\alpha}$ in Eq. (68) by $R^{-\alpha} + \epsilon R^\alpha$; then the exponent in integral $I^+(x)$ acquires $x(z + \epsilon/z)$ and a saddle point $z_c = \sqrt{\epsilon}$ appears, through which contour $C$ should be drawn. The integration should be carried out in the upward direction so that contour $C$ is obtained from the initial contour $C_0$ by rotation through $\pi/2$ in the positive direction around point $z_c$. For $\epsilon \longrightarrow 0$, the saddle point disappears, but the direction of contour $C$ is preserved and corresponds to the negative sign in expression (75). Evaluating integral $I^+(x)$ using the deformation of contour $C$ to position $C_1$, we obtain

$$I^+(x) \longrightarrow \frac{2\pi i}{\Gamma(1 + \nu)} x^\nu, \quad (76)$$

which gives the following expression for the disconti-nuity at the cut of function $G(g)$:

$$\Delta G(g) = 2\pi i \frac{BA^+}{\Gamma(1+\nu)}\left(\frac{2S_0}{g}\right)^{M/2+r+\nu}\exp\left(-\frac{2S_0}{g}\right). \quad (77)$$

The Lipatov Asymptotics for $G_N$ is obtained in accordance with the correspondence rule (30),

$$G_N = \frac{BA^+}{\Gamma(1+\nu)}(2S_0)^{-N}\Gamma(N+M/2+r+\nu), \quad (78)$$

and is well defined for all values of $\nu > -1$. Expression (76) corresponds to the algorithm correctly formulated by Bogomolny and Fateyev [9], although the line of their reasoning leads to the result with the opposite sign.[13]

**Asymptotic form of coefficients $A_K$.** In analogy with expression (42), we have the following exact expression for coefficients $A_K$ corresponding to quantity $G(g)$:

$$A_K = \frac{1}{\sqrt{2\pi}c}\int\frac{d\epsilon}{2\pi i\epsilon}\int\frac{dg}{2\pi ig}\epsilon^{(M+r-1)/2}G(g)$$

$$\times \exp\left\{\frac{1}{\epsilon}\left(\frac{S_0\epsilon}{g}+1\right)-K\ln\epsilon\right\}. \quad (79)$$

Substituting quantity $G^{(2)}(g)$ from Eq. (71) for $G(g)$ and evaluating the integrals with respect to $g$ and $\epsilon$ in the saddle-point approximation, we obtain

$$A_K = \pm\frac{B}{c(2\pi)^2(\ln 2)^{r/2}}$$

$$\times \Gamma\left(K+\frac{r}{2}\right)(\ln 2)^{-K}F\left(\frac{K}{\ln 2}\right). \quad (80)$$

Vertical contours of integration with respect to $g$ and $\epsilon$ correspond to "poor" Gaussian integrals (cf. formula (44)) and must be rotated simultaneously to the horizontal position in accordance with Section 3. The sign in formula (80) turns out to be indeterminate for the same reason as in the above case. To fix the sign, we replace $R^{-\alpha}$ by $R^{-\alpha}+\epsilon R^{\alpha}$ and consider the saddle-point configuration in the range of parameters, where $S_{int} > 0$, which corresponds to integral $I^-(x)$. For this region, $\det S''\{\psi_c\} > 0$ and the sign in formula (80) should correspond to the arithmetic root of the determinant, which leads to the condition $\pm B > 0$. Formula (80) is valid if function $F(x)$ is well defined on the real axis. Otherwise, $F(x)$ should be treated as the half-sum of $F(x+i0)$ and $F(x-i0)$ since half the saddle point contribution stems from region $\text{Im}\,g > 0$, while the other half, from region $\text{Im}\,g < 0$. Thus, we can write

$$A_K = \frac{|B|}{c(2\pi)^2(\ln 2)^{r/2}}\Gamma\left(K+\frac{r}{2}\right)$$

$$\times(\ln 2)^{-K}\text{Re}\,F\left(\frac{K}{\ln 2}+i0\right) \quad (81)$$

and the substitution of Eq. (74) gives

$$A_K = \frac{|B|\Gamma(-\nu)(A^- + A^+\cos\pi\nu)}{c(2\pi)^r(\ln 2)^{\nu+r/2}}$$

$$\times\Gamma\left(K+\frac{r}{2}+\nu\right)(\ln 2)^{-K}. \quad (82)$$

_____________

[13]The incorrect sign also appeared in the work by Balitsky [22] who calculated the Lipatov asymptotics in QCD and was corrected in the review by Zakharov [23] proceeding from physical considerations.

The poorly defined integral $I^+(x)$ is treated as the half-sum of integrals over contours $C'$ an $C''$ (see Fig. 3c),

$$I^+(x) \longrightarrow \frac{1}{2}[I^+(x+i0)+I^+(x-i0)]$$

$$= \Gamma(-\nu)x^\nu\cos\pi\nu \quad (83)$$

(cf. formula (76)). If the saddle point is created artificially by replacing $R^{-\alpha}$ by $R^{-\alpha}+\epsilon R^{\alpha}$, contours $C'$ and $C''$ correspond to the steepest decent as before and the saddle point $z_c$ makes zero contribution to $A_K$ in accordance with the fact that $\det S''\{\psi_c\} < 0$ for this point (Section 4). It can be seen that for a power law of instanton interaction (68), the argument of the gamma-function acquires an additional contribution $\nu = d/\alpha$ as compared to formula (4). This contribution is in fact important only for $d = 4$. For $d < 4$, the interaction of instantons is exponential, which corresponds to $\alpha = \infty$ and $\nu = 0$: the argument of the gamma-function corresponds to Eq. (4), but additional logarithmic factors appear (Section 7).

Interpretation of integral $I^+(x)$ completes the formulation of the general computational algorithm. Proceeding from the known expression (51) for the one-instanton contribution, we construct the contribution from two-instanton configuration (64), in which we should calculate the interaction of instantons $S_{int}(\lambda,\lambda')$ in the region where this interaction is weak. Introducing the slowly varying function $F(x)$ in accordance with formula (71), we can straightaway write the result (81).

## 7. COEFFICIENTS $A_K$ IN THE $\varphi^4$ THEORY

### 7.1. Interaction of Instantons

Action (5) is reduced to the form (15) using the substitution[14] $\varphi_\alpha(x) = (-g)^{-1/2}\phi_\alpha(x)$. An instanton is pre-sented in the form $\phi_\alpha(x) = u_\alpha \phi_c(x)$, where $u_\alpha$ are the components of the unit vector $\mathbf{u}$ and $\phi_c(x)$ is the solution to the equation

$$-\Delta\phi + m^2\phi - \phi^3 = 0. \qquad (84)$$

_____________

[14]The additional minus as compared to relation (15) ensures the real-valuedness of $\phi_c$. In this case, the substitutions $g^{-b} \longrightarrow (-g)^{-b}$ and $S_0^{-N} \longrightarrow (-S_0)^{-N}$ take place in Eqs. (22) and (24).

Using the general form of the two-instanton configuration,

$$\psi_\alpha(x) = u_\alpha\phi_c(x) + u'_\alpha\phi_c(x+R) \equiv u_\alpha\phi + u'_\alpha\phi_R, \quad (85)$$

and definition (52) for the interaction of instantons, we have

$$S_{\text{int}}\{u, u', R\} = \int d^t x$$
$$\times \left\{ (uu')\phi^3\phi_R + (uu')^2\phi^2\phi_R^2 + \frac{1}{2}\phi^2\phi_R^2 \right\}, \qquad (86)$$

where we have used definition (52) and eliminated gradients with the help of Eq. (84). For $d < 4$, we can assume that $m = 1$, since the substitution $x \longrightarrow x/m$, $\phi \longrightarrow \phi m$ eliminates the mass from Eq. (84), after which it disappears in all dimensionless quantities. For large values of $|x|$, the spherically symmetric solution to Eq. (84) has the form

$$\phi_c(x) = \text{const}|x|^{-\mu}K_\mu(|x|)$$
$$= \text{const}|x|^{(1-d)/2}e^{-|x|}, \qquad (87)$$

where $\mu = (d-2)/2$ and $K_\mu(x)$ is the Macdonald function. The substitution of relation (87) into Eq. (86) shows that the main contribution is determined by the first term,

$$S_{\text{int}}\{u, u', R\} \approx (\mathbf{u} \cdot \mathbf{u}')\phi_c(R + \bar{x})\int\phi_c^3(x)d^t x$$
$$\approx \text{const}(\mathbf{u} \cdot \mathbf{u}')R^{(1-d)/2}e^{-R}, \qquad (88)$$

where $\bar{x} \sim 1$. For $d = 1$, the constant can be evaluated using the explicit form of instanton $\phi_c(x) = \sqrt{2}/\cosh x$,

$$S_{\text{int}}\{u, u', R\} = 16(\mathbf{u} \cdot \mathbf{u}')e^{-R}. \qquad (89)$$

In the four-dimensional case, the massless theory is of principal importance, in which Eq. (84) has a solution

$$\phi_c(x) = \frac{2\sqrt{2}\rho}{x^2 + \rho^2} \qquad (90)$$

with an arbitrary instanton radius $\rho$. In the linear combination (85), we must permit various radii $\rho$ and $\rho_1$

for functions $\phi$ and $\phi_R$. Disregarding quantity $\bar{x}$ in relation (88), we obtain

$$S_{\text{int}}\{\mathbf{u}, \mathbf{u}', \rho, \rho_1, R\} = 32\pi^2(\mathbf{u} \cdot \mathbf{u}')\frac{\rho\rho_1}{R^2 + \rho_1^2}, \quad (91)$$

where we assume that

$$\rho \ll \rho_1 \sim R, \qquad (92)$$

since precisely these configurations are of interest for the subsequent analysis.[15]

### 7.2. Results for $d < 4$.

Proceeding from the well-known Lipatov asymptotics (formula (79) in [18]) and using the correspondence rule (30), for the one-instanton contribution, we obtain

$$Z^{(1)}(\alpha_1 x_1, ..., \alpha_M x_M)$$
$$= ic_0(-g)^{-(M+r)/2}\exp(-S_0/g)$$
$$\times \int d^t x_0\phi_c(x_1 - x_0)...\phi_c(x_M - x_0) \qquad (93)$$
$$\times \int d^t u\delta(|u| - 1)u_{\alpha_1}...u_{\alpha_M},$$

where

$$S_0 = -\frac{I_4}{4}, \quad r = n - 1 + d, \quad I_p = \int d^t x\phi_c^p(x), \quad (98b)$$

$$c_0 = \frac{1}{(2\pi)^{r/2}}\left(\frac{I_6 - I_4}{d}\right)^{d/2}$$
$$\times I_4^{(n-1)/2}\left[-\bar{D}_R(1)\bar{D}_R^{n-1}\left(\frac{1}{3}\right)\right]^{-1/2} \qquad (94)$$

and $\phi_c(x)$ is the solution to Eq. (84) with $m = 1$; $\bar{D}_R(1)$ and $\bar{D}_R(1/3)$ are the renormalized determinants whose values will be given below. In accordance with the instanton combination rule, we write the expression for the two-instanton contribution

$$G_M^{(2)}(\alpha_1 x_1, ..., \alpha_M x_M) = -c_0^2(-g)^{-(M+2r)/2}$$
$$\times \exp\left(-\frac{2S_0}{g}\right)H(g)\int d^t x_0\phi_c(x_1 - x_0)...\phi_c(x_M - x_0) \quad (98b)$$
$$\times \int d^t u\delta(|\mathbf{u}| - 1)u_{\alpha_1}...u_{\alpha_M},$$

---

[15]The possibility of disregarding the cross terms in Eq. (53) in this case is related not with the large distance between the instantons, but with different degrees of localization for $\phi_\lambda$ and $\phi_{\lambda'}$.

$$H(g) = \int d^n u' \delta(|\mathbf{u}| - 1)$$

$$\times \int d^t x_0' \left\{ \exp\left[ -\frac{S_{\text{int}}(\mathbf{u}, \mathbf{u}', x_0' - x_0)}{g} \right] - 1 \right\}. \quad (95)$$

The last integral ($R \equiv x_0' - x_0$),

$$\int d^t R \left( \exp\left\{ -\frac{A R^{(1-d)/2} e^{-R}}{g} \right\} - 1 \right) \approx -\frac{\sigma_d}{d}\left( \ln\left| \frac{1}{g} \right| \right)^d, \quad (96)$$

is independent of $A$ to a logarithmic accuracy ($\sigma_d$ is the area of a unit sphere in the $d$-dimensional space); for this reason, the dependence on $\mathbf{u}'$ is absent and $H(g)$ can be obtained from relation (96) simply by multiplying it by $\sigma_n$. Following the algorithm described in Section 6, we obtain the following result for $A_K$:

$$A_K = -c_0 \frac{2^{-M/2}}{2\pi}(I_4 \ln 2)^{-r/2}\frac{\sigma_n \sigma_d}{d}$$

$$\times \Gamma\left( K + \frac{\Gamma}{2} \right)(\ln 2)^{-K} \ln^d K. \quad (97)$$

Using the numerical values [21]

$$I_6 = 71.080, \quad I_4 = 23.402,$$

$$-\overline{D}_R(1) = 135.3, \quad \overline{D}(1/3) = 1.465 \quad (98a)$$

for $d = 2$ and

$$I_6 = 659.87, \quad I_4 = 75.589,$$

$$-\overline{D}_R(1) = 10.544, \quad \overline{D}(1/3) = 1.4571 \quad (98b)$$

for $d = 3$, we obtain formulas (9) and (10) given in the Introduction. For $d = 1$, the result can be obtained not with a logarithm, but with a power accuracy in $1/K$. Integral (96) for $\mathbf{u} \cdot \mathbf{u}'/g > 0$ has the form

$$\sigma_1 \int_0^\infty dR \left( \exp\left\{ -\frac{16(\mathbf{u} \cdot \mathbf{u}')e^{-R}}{g} \right\} - 1 \right)$$

$$= -2\left( \ln\frac{16(\mathbf{u} \cdot \mathbf{u}')}{g} + C_E \right) \quad (99)$$

and integration with respect to $\mathbf{u}'$ gives

$$\text{Re}\, H(g + i0)$$

$$= -2\sigma_n \left( \ln\frac{16}{g} + C_E + \frac{\psi(1/2) - \psi(n/2)}{2} \right). \quad (100)$$

Passing to the expression for $A_K$ and using the values of

parameters [4]

$$I_4 = \frac{16}{3}, \quad I_6 = \frac{128}{15},$$

$$\overline{D}_R(1) = -\frac{1}{5}, \quad \overline{D}_R\left( \frac{1}{3} \right) = \frac{1}{3}, \quad (101)$$

we obtain result (8).

### 7.3. Four-Dimensional Case

In the four-dimensional case, the expression for the one-instanton contribution differs from (93) in view of the presence of an extra zero mode associated with the possibility of variation of instanton radius $\rho$. This expression can be derived from formula (113) from [19] by using the correspondence rule (30):

$$Z_M^{(1)}(\alpha_1 x_1, \ldots, \alpha_M x_M) = i c_0 (-g)^{-(M+r)/2} \exp(-S_0/g)$$

$$\times \int d^4 x_0 \int \frac{d\rho}{\rho^{M+5}} \exp(\nu \ln \mu \rho) \quad (102)$$

$$\times \phi_c\left( \frac{x_1 - x_0}{\rho} \right) \ldots \phi_c\left( \frac{x_M - x_0}{\rho} \right) \int d^n u\, \delta(|\mathbf{u}| - 1) u_{\alpha_1} \ldots u_{\alpha_M},$$

where $\phi_c(x)$ is the instanton solution (90) with $\rho = 1$, $\mu$ is the momentum of charge normalization,[16]

$$c_0 = \frac{1}{(2\pi)^{r/2}}\left( \frac{I_6}{4} \right)^2 J^{1/2} I_4^{(n-1)/2}\left[ -\overline{D}_R(1)\overline{D}_R^{n-1}\left( \frac{1}{3} \right) \right]^{-1/2}$$

$$\times \exp\left[ -\frac{3}{4}r + \nu\left( -\frac{1}{2}\ln 3 + C_E - \frac{1}{6} \right) \right], \quad (103)$$

$$r = n + 4, \quad \nu = (n+8)/3,$$

$$J = \int d^4 x\, 3\phi_c^2(x)[\partial\phi_c(x)/\partial\rho]_{\rho=1}^2$$

and expressions for $S_0$ and $I_p$ are the same as in (94). The two-instanton contribution has the form

$$G_M^{(2)}(\alpha_1 x_1, \ldots, \alpha_M x_M) = -c_0^2(-g)^{-(M+2r)/2}\exp\left( -\frac{2S_0}{g} \right)$$

$$\times \int d^n u\, \delta(|\mathbf{u}| - 1) u_{\alpha_1} \ldots u_{\alpha_M} \int \frac{d\rho}{\rho^{M+5}}\exp(\nu \ln \mu \rho)$$

$$\times \int d^4 x_0\, \phi_c\left( \frac{x_1 - x_0}{\rho} \right) \ldots \phi_c\left( \frac{x_\mu - x_0}{\rho} \right) \quad (104)$$

$$\times \int d^n u'\, \delta(|\mathbf{u}'| - 1) \int \frac{d\rho_1}{\rho_1^5}\exp(\nu \ln \mu \rho_1)$$

---

[16]For a transition from the bare 1 to a renormalized charge in the formulas from [19], we must carry out the substitution $\ln \Lambda \rho - \ln 2 + C_E + 1/3 \longrightarrow \ln \mu \rho - (1/2)\ln 3 + C_E - 1/6$.

$$\times \int d^4 x_0' \left\{ \exp\left[-\frac{S_{\text{int}}(u, u', \rho, \rho_1, x_0' - x_0)}{g}\right] - 1 \right\}.$$

Using the relation[17]

$$\int_0^\infty \frac{d\rho_1}{\rho_1^5} \exp(\nu \ln \mu \rho_1) \int d^4 R\, \delta\left(z - \frac{\rho \rho_1}{R^2 + \rho_1^2}\right)$$

$$= \frac{\pi^2 \exp(\nu \ln \mu \rho)}{(\nu - 1)(\nu - 2) z^{1+\nu}} \tag{105}$$

we can transform the integral with respect to $\rho_1$ and $x_0'$ in expression (104) to the form

$$\frac{\pi^2 \exp(\nu \ln \mu \rho)}{(\nu - 1)(\nu - 2)}$$

$$\times \int_0^\infty \frac{dz}{z^{1+\nu}} \exp\left(-\frac{32\pi^2(\mathbf{u} \cdot \mathbf{u}')}{g} z\right) \tag{106}$$

and obtain an integral of the type (73). Following the algorithm described in Section 6, for the asymptotic form of $A_K$ we obtain

$$A_K = c_0 \frac{2^{-M/2}}{(I_4 \ln 2)^{r/2}} \frac{\pi}{4} \left(\frac{64\pi^2 K}{I_4 \ln 2}\right)^\nu \sigma_n \frac{\Gamma\left(\dfrac{n}{2}\right) \Gamma\left(\dfrac{1+\nu}{2}\right)}{\sqrt{\pi} \Gamma\left(\dfrac{n+\nu}{2}\right)} \tag{107}$$

$$\times \frac{\Gamma(-\nu)(1 + \cos \nu \pi)}{(\nu - 1)(\nu - 2)} Q \Gamma\left(K + \frac{r}{2}\right)(\ln 2)^{-K},$$

where the quantity

$$Q = \int \frac{d\rho}{\rho^{M+5}} \exp(2\nu \ln \mu \rho)$$

$$\times \int d^l x_0 \phi_c\left(\frac{x_1 - x_0}{\rho}\right) \ldots \phi_c\left(\frac{x_M - x_0}{\rho}\right)$$

$$\times \left[\int \frac{d\rho}{\rho^{M+5}} \exp(\nu \ln \mu \rho)\right.$$

$$\left. \times \int d^l x_0 \phi_c\left(\frac{x_1 - x_0}{\rho}\right) \ldots \phi_c\left(\frac{x_M - x_0}{\rho}\right)\right]^{-1} \tag{108}$$

substantially depends on the external coordinates $x_1$, ..., $x_M$. The expression for this quantity can be slightly simplified by passing to the momentum space and by choosing the values of external momenta $p_i$ on the order of $p$, estimating their values at a symmetric point ($\mathbf{p}_i \cdot \mathbf{p}_j = p^2(4\delta_{ij} - 1)/3$),

$$Q = \frac{\displaystyle\int_0^\infty dy\, y^{2M-5+2\nu} K_1^M(y)}{\displaystyle\int_0^\infty dy\, y^{2M-5+\nu} K_1^M(y)} \exp\left(\frac{\nu \ln \mu}{p}\right), \tag{109}$$

where $K_1(y)$ is the Macdonald function. Substituting numerical values

$$I_4 = \frac{32}{3}\pi^2, \quad I_6 = \frac{128}{5}\pi^2, \quad J = \frac{32}{15}\pi^2,$$

$$\overline{D}_R(1) = -578, \quad \overline{D}_R(1/3) = 0.872 \tag{110}$$

leads to

$$A_K = \frac{2^{-M/2}}{20.2} 0.842^n \frac{\Gamma(-\nu)(1 + \cos \pi \nu)}{(n+2)(n+5)}$$

$$\times \frac{\Gamma\left(\dfrac{1+\nu}{2}\right)}{\Gamma\left(\dfrac{n+\nu}{2}\right)} Q \Gamma\left(K + \frac{r}{2} + \nu\right)(\ln 2)^{-K} \tag{111}$$

and to the numerical results given in the Introduction.

# 8. ASYMMETRIC INSTANTON

In this section, we consider the situation when the second instanton is well localized in space, but does not possess any special symmetry. In view of the absence of soft modes, we can use the algorithm developed in Section 4, according to which it is sufficient to calculate the contribution of an asymmetric instanton to the Lipatov integral (18). Following the line of reasoning in Section 7, we assume that

$$\varphi_\alpha(x) = (-g)^{-1/2} \phi_\alpha(x), \quad \phi_\alpha(x) = u_\alpha \phi_c(x), \tag{112}$$

and divide fluctuations into longitudinal and transverse:

$$\delta\varphi_\alpha(x) = \delta\varphi^L(x) u_\alpha + \delta\varphi_\alpha^T(x),$$

$$\sum_\alpha \delta\varphi_\alpha^T(x) u_\alpha = 0. \tag{113}$$

---

[17]In deriving formula (105), it is found that the main contribution to the integral appears from region $\rho_1 \sim R$. In the subsequent analysis, the values of $\rho_1$ and $R$ are large in view of the large value of parameter $1/g$, while the value of $\rho$ turns out to be on the order of the reciprocal external momentum. This justifies condition (92).

In analogy with relation (20), we then obtain

$$Z_N = e^{-N}(-g_c)^{-M/2}g_c^{-N}\int_{-\infty}^{\infty}\frac{dt}{2\pi}\int D\varphi^L$$

$$\times \int D\varphi_\alpha^T u_{\alpha_1}\ldots u_{\alpha_M}\phi_c(x_1)\ldots\phi_c(x_M) \qquad (114)$$

$$\times \exp\left\{\frac{Nt^2}{2} - \frac{1}{2}(\delta\varphi^L, \hat{M}_L\delta\varphi^L) - \frac{1}{2}\sum_{\alpha=1}^{n-1}(\delta\varphi_\alpha^T, \hat{M}_T\delta\varphi_\alpha^T)\right\},$$

where

$$\begin{aligned}\hat{M}_L &= \hat{p}^2 + m^2 - 3\phi_c^2(x),\\ \hat{M}_T &= \hat{p}^2 + m^2 - \phi_c^2(x).\end{aligned} \qquad (115)$$

The integrals with respect to longitudinal and transverse fluctuations are factorized so that zero modes for operators $\hat{M}_L$ and $\hat{M}_T$ can be singled out independently. Carrying out the orthogonal transformation of variables $\delta\varphi$, which diagonalizes $\hat{M}_L$ and $\hat{M}_T$, we perform the division in accordance with the scheme

$$\delta\varphi = \delta\varphi' + \delta\tilde{\varphi}, \quad D\varphi = D(\delta\varphi')D(\delta\tilde{\varphi}) \qquad (116)$$

for $\delta\varphi^L$ and $\delta\varphi_\alpha^T$, where tildes and primes mark the subspace of zero modes and the space orthogonal to it, respectively. Performing Gaussian integration over nonzero modes, we obtain

$$Z_N = \frac{(-S\{\phi_c\})^{-(M+r)/2}}{(2\pi)^{1+r/2}}S\{\phi_c\}^{-N}\Gamma\left(N+\frac{M+r}{2}\right)$$

$$\times \sqrt{\frac{D_0}{D_L'}\left(\frac{D_0}{D_T'}\right)^{n-1}}\int D(\delta\tilde{\varphi}^L)\phi_c(x_1)\ldots\phi_c(x_M) \qquad (117)$$

$$\times \int D(\delta\tilde{\varphi}_\alpha^T)u_{\alpha_1}\ldots u_{\alpha_M},$$

where $D_0 = \det S''\{0\}$, $D_L' = \det[\hat{M}_L]_{P'}$, and $D_T' = \det[\hat{M}_T]_{P'}$.

It is well known that the existence of zero modes is associated with the symmetry of action, $S\{\phi\} = S\{\hat{L}\phi\}$, relative to a certain continuous group of transformations defined by operator $\hat{L}$; if $\phi_c$ is an instanton (i.e., the solution to the equation $S'(\phi_c) = 0$), $\hat{L}\phi_c$ is also an instanton ($S'(\hat{L}\phi_c) = 0$). Using the infinitesimal form of operator $\hat{L}$ close to the unit operator, $\hat{L}_\epsilon = 1 + \epsilon\hat{T}$, we can easily see that $\hat{T}\phi_c$ is a zero mode of operator $S''\{\phi_c\}$, which is obviously connected with the genera-

tor of group $\hat{T}$. In the given case, the following groups of transformations are significant:

(a) rotations in the vector space,

$$\hat{L}^T\phi_\alpha(x) = g_{\alpha\beta}\phi_\beta(x), \qquad (118)$$

where $g_{\alpha\beta}$ are the elements of an orthogonal matrix;

(b) translations

$$\hat{L}^t(x_0)\phi(x) = \phi(x + x_0); \qquad (119)$$

(c) dilatation for $d = 4$, associated with the scale invariance of the massless four-dimensional theory,

$$\hat{L}^\partial(\ln\rho)\phi(x) = \rho\phi(\rho x); \qquad (120)$$

(d) rotations in the coordinate space,

$$\hat{L}^r\{\theta_s\}\phi(x) = \phi(\hat{g}x), \qquad (121)$$

where $\hat{g} = \hat{g}\{\theta_s\}$ is the orthogonal matrix defined by the rotational angles $\theta_s$.

Transformation (118) is reduced to the rotation of the unit vector $\mathbf{u}$ in (112) and generates the zero mode

$$h_T(x) = \phi_c(x) \qquad (122)$$

of operator $\hat{M}_T$; in expression (114), this mode is $(n - 1)$-fold degenerate. The separation of this mode is performed in the conventional way [24] and corresponds to the following substitution in formula (117):

$$\int D(\delta\tilde{\varphi}_T)u_{\alpha_1}\ldots u_{\alpha_M}$$

$$\longrightarrow I_2^{(n-1)/2}\int d^n u\,\delta(|u|-1)u_{\alpha_1}\ldots u_{\alpha_M}, \qquad (123)$$

where integrals $I_p$ are defined in (94). Infinitesimal forms of operators $L^t$, $L^\partial$, and $L^r$ can be written as

$$L^t(\delta x_0) = 1 + \sum_i \delta x_{0,i}\frac{\partial}{\partial x_i} \equiv 1 + \sum_i \delta x_{0,i}\hat{T}_i^t,$$

$$L^\partial(\epsilon) = 1 + \epsilon\left(1 + \sum_i x_i\frac{\partial}{\partial x_i}\right) \equiv 1 + \epsilon\hat{T}^\partial,$$

$$\qquad (124)$$

$$L^r(\delta\theta_s) = 1 + \sum_s \delta\theta_s\hat{T}_s^r,$$

$$\hat{T}_s^r \equiv \hat{T}_{(ij)}^r = x_i\frac{\partial}{\partial x_j} - x_j\frac{\partial}{\partial x_i},$$

where each operator $\hat{T}_s^r$ corresponds to rotation in one of the $d(d-1)/2$ planes $(x_i, x_j)$ and subscript $s$ labels these planes. Accordingly, the following zero modes belonging to operator $\hat{M}_L$ exist:

$$h_i^t(x) = \frac{\partial \phi_c(x)}{\partial x_i},$$

$$h^{\partial}(x) = \phi_c(x) + \sum_i x_i \frac{\partial \phi_c(x)}{\partial x_i}, \qquad (125)$$

$$h_s^r(x) \equiv h_{(ij)}^r(x) = x_i \frac{\partial \phi_c(x)}{\partial x_j} - x_j \frac{\partial \phi_c(x)}{\partial x_i}.$$

The existence of rotational modes $h_s^r(x)$ is a specific feature of a spherically asymmetric instanton; these modes have not been considered earlier. The nontrivial moments are connected with non-Abelian nature of the group of transformations and with nonorthogonality of the basis constructed from vectors (125).

The complete group of transformations is determined by the operator

$$\hat{L}\{\theta_s, \ln\rho, x_0\} = \hat{L}^r\{\theta_s\}\hat{L}^{\partial}\{\ln\rho\}\hat{L}^t\{x_0\}, \qquad (126)$$

so that

$$\hat{L}f(x) = \rho f(\hat{g}\rho(x + x_0)). \qquad (127)$$

The infinitesimal form of this operator,

$$\hat{L}(\delta\mu_i) = 1 + \sum_i \delta\mu_i \hat{T}_i \qquad (128)$$

includes as $\hat{T}_i$ all generators $\hat{T}_i^t$, $\hat{T}^{\partial}$, and $\hat{T}_s^r$ introduced by relations (124), while $\mu_i$ labels the variables $x_{0,i}$, $\ln\rho$, and $\theta_s$.

Following the algorithm used in Section 2, we introduce the expansion of unity into the integrand in expression (114) ($r_L$ is the number of zero modes of operator $\hat{M}_L$),

$$1 = \prod_{i=1}^{r_L} \int d\lambda_i \delta\left(\lambda_i - \frac{\int d^l x\, \phi^4(x) f^{(i)}(x)}{\int d^l x\, \phi^4(x)}\right)$$

$$= (I_4)^{r_L} \prod_{i=1}^{r_L} \int d\lambda_i \delta\left(\lambda_i \int d^l x\, 4\phi_c^3(x)\delta\phi(x) \right. \qquad (129)$$

$$\left. - \int d^l x\, 4\phi_c^3(x)\delta\phi(x) f^{(i)}(x)\right),$$

where we specified to a certain extent the form of functionals $f_i(\phi)$ in expression (21) by introducing the coor-dinate functions $f^{(i)}(x)$,[18] carried out the expansion in the vicinity of the saddle-point configuration, and chose the instanton from the condition

$$\lambda_i = \frac{\int d^l x\, \phi_c^4(x) f^{(i)}(x)}{\int d^l x\, \phi_c^4(x)}. \qquad (130)$$

To make clear the arbitrariness in the choice of the instanton, we represent $\phi_c(x)$ as the result of action of operator $\hat{L}^{-1}$ on a certain fixed function $\bar{\phi}_c(x)$,

$$\phi_c(x) = \hat{L}^{-1}\bar{\phi}_c(x) = \frac{1}{\rho}\bar{\phi}_c\left(\frac{\hat{g}^{-1}x}{\rho} - x_0\right). \qquad (131)$$

In this case, we obtain

$$\lambda_i \equiv \frac{\int d^l x\, \bar{\phi}_c^4(x)\rho^{-1}\hat{L}\{\theta_s, \ln\rho, x_0\} f(x)}{\int d^l x\, \bar{\phi}_c^4(x)}$$

$$\equiv F^{(i)}\{\theta_s, \ln\rho, x_0\}. \qquad (132)$$

Let us expand $\delta\phi(x)$ in orthonormal eigenfunctions $e_j(x)$ of operator $\hat{M}_L$:

$$\delta\phi(x) = \sum_{j=1}^{r_L} C_j e_j(x) + \ldots = \sum_{j=1}^{r_L} B_j h_j(x) + \ldots, \qquad (133)$$

here, we have singled out the terms corresponding to the subspace of zero modes and reexpanded these terms in functions (125). Integration of $D(\delta\tilde{\phi}^L)$ in expression (117) is in fact the integration with respect to coefficients $C_j$:

$$\int D(\delta\tilde{\phi}_L) \longrightarrow \int \prod_{j=1}^{r_L} dC_j = \int (\det\Gamma)^{1/2} \prod_{j=1}^{r_L} dB_j, \qquad (134)$$

where $\Gamma$ is the Gram matrix plotted on vectors (125). Substituting expression (133) into (129) and taking into account that

$$\int d^l x\, 4\phi_c^3(x) h_j(x) = 0 \qquad (135)$$

---

[18] In fact, the results are independent of this choice, which is manifested in the fact that functions $f^{(i)}(x)$ do not appear in the final formula (141).

for all zero modes $h_j(x)$, we obtain

$$1 = (I_4)^{r_L} \prod_i \int d\lambda_i$$

$$\times \delta\left(-\sum_j B_j \int d^l x \, 4\phi_c^3(x) f^{(i)}(x) h_j(x) + \ldots\right). \quad (136)$$

Now, we can easily prove that

$$\int d^l x \, 4\phi_c^3(x) h_j(x) f^{(i)}(x)$$

$$= -\int (d^l x) \phi_c^4(x) \hat{T}_j f^{(i)}(x) \quad (137)$$

for translations and rotations, while an analogous formula with $\hat{T}_j \longrightarrow \hat{T}_j - 1$ is valid for dilatation. We perform the variation of variables $\theta_s \longrightarrow \theta_s + \delta\theta_s$, $\ln\rho \longrightarrow \ln\rho + \varepsilon$, and $x_0 \longrightarrow x_0 + \delta x_0$ in Eqs. (132) and take into account the group relation

$$\hat{L}\{\theta_s + \delta\theta_s, \ln\rho + \varepsilon, x_0 + \delta x_0\}$$

$$= \hat{L}\{\delta\theta_s', \varepsilon', \delta x_0'\} \hat{L}\{\theta_s, \ln\rho, x_0\}, \quad (138)$$

where the primed and nonprimed increments do not coincide (in view of the non-Abelian nature of the group), but are connected via a linear transformation. It can easily be proved that $\varepsilon' = \varepsilon$, $\delta x_0' = \hat{g}\rho\delta x_0$, and the relation between $\delta\theta_s'$ and $\delta\theta_s$ is defined by the relation [25]

$$\hat{g}\{\theta_s + \delta\theta_s\} = \hat{g}\{\delta\theta_s'\}\hat{g}\{\theta_s\}$$

$$\text{or} \quad \delta\theta_s = \sum_{s'} J_{ss'}\{\theta_s\}\delta\theta_{s'}'. \quad (139)$$

Using the infinitesimal form of operator $\hat{L}\{\delta\theta_s', \varepsilon', \delta x_0'\}$, we obtain from relations (132)

$$\frac{\int d^l x \phi_c^4 \hat{T}_j' f^{(i)}(x)}{\int d^l x \phi_c^4 x} = \frac{1}{\rho}\sum_l g_{jl}\frac{\partial F^{(i)}}{\partial x_{0,l}},$$

$$\frac{\int d^l x \phi_c^4(\hat{T}^\partial - 1) f^{(i)}(x)}{\int d^l x \phi_c^4 x} = \frac{\partial F^{(i)}}{\partial \ln\rho}, \quad (140)$$

$$\frac{\int d^l x \phi_c^4 \hat{T}_s' f^{(i)}(x)}{\int d^l x \phi_c^4 x} = \sum_{s'} J_{s's}\{\theta_s\}\frac{\partial F^{(i)}}{\partial \theta_{s'}}.$$

Substituting Eqs. (137) and (140) into expression (136) and introducing the partition of unity obtained in this

way into relation (134), we obtain

$$\int \prod_j dC_j = \int \prod_i d\lambda_i \frac{\rho^d (\det\Gamma)^{1/2}}{\det\|\partial F^{(i)}/\partial\mu_j\|\det J\{\theta_s\}}$$

$$= \int \prod_j d\mu_j \frac{\rho^d (\det\Gamma)^{1/2}}{\det J\{\theta_s\}} = \int \frac{\rho^d (\det\Gamma)^{1/2}}{\det J\{\theta_s\}} d^l x_0 d\ln\rho \quad (141)$$

$$\times \prod_s d\theta_s = \int \rho^d (\det\Gamma)^{1/2} d^l x_0 d\ln\rho \, d\tau_g,$$

We used the fact that the quantity $\prod_s d\theta_d/\det J\{\theta_s\}$ is the definition of the invariant measure of integration $d\tau_g$ over the group of rotations [25]. Using relation (131) and performing the substitution $x_{0, i} \longrightarrow x_{0, i}/\rho$, we obtain the sought rule of transition to collective variables,

$$\int D(\delta\tilde{\phi}_L)\phi_c(x_1)\ldots\phi_c(x_M)$$

$$\longrightarrow \int (\det\Gamma)^{1/2}(d^l x_0) d\ln\rho \, d\tau_g \rho^{-M} \quad (142)$$

$$\times \bar{\phi}_c\left(\frac{\hat{g}^{-1}(x_1 - x_0)}{\rho}\right)\ldots\bar{\phi}_c\left(\frac{\hat{g}^{-1}(x_M - x_0)}{\rho}\right),$$

which is valid for $d = 4$. The result for $d < 4$ can be obtained by setting $\rho = 1$ and eliminating the integration with respect to $\ln\rho$. The expression for invariant measure $d\tau_g$ depends on the method for parametrizing matrix $\hat{g}$; if we use the Euler angles $\theta_l^k$, this expression has the form

$$d\tau_g = \prod_{k=1}^{d-1}\prod_{l=1}^{k}\sin^{l-1}\theta_l^k d\theta_l^k,$$

$$0 \leq \theta_1^k < 2\pi, \quad 0 \leq \theta_l^k < \pi \ (l \neq 1). \quad (143)$$

When the Euler angles are introduced in the $d$-dimensional space, the rotation matrix can be written in the form [26]

$$\hat{g} = \hat{g}^{(d-1)}\ldots\hat{g}^{(2)}\hat{g}^{(1)}, \quad \hat{g}^{(k)} = \hat{g}_1(\theta_1^k)\hat{g}_2(\theta_2^k)\ldots\hat{g}_k(\theta_k^k),$$

where $\hat{g}_i(\theta) \equiv \hat{g}_{i+1,i}(\theta)$ and $\hat{g}_{ij}(\theta)$ is the matrix of rotation through angle $\theta$ in the plane $(x_i, x_j)$. Substitution (131) is also carried out in the Gram matrix; as a result, its determinant turns out to be a function of collective variables. In fact, the dependence on $\rho$ can be factorized, $(\det\Gamma)^{1/2} = \rho^{-4}(\det\bar{\Gamma})^{1/2}$ (for $d = 4$), and the dependence on $x_0$ is ruled out in view of the possibility of transition to linear combinations of rows and col-umns in the determinant; apparently the dependence on the angles of rotation is also absent.[19]

The nonorthogonality of zero modes $h_i(x)$ is also taken into account in the transformations of determinants according to Brézin and Parisi (see notation in [4, 19]),

$$\frac{D'_L}{D_0} = \frac{\det \Gamma}{\det G} \overline{D}(1), \quad D(1) = \lim_{z \to 1} \frac{D(z)}{(1-z)^{r_L}},$$

$$\frac{D'_T}{D_0} = \frac{I_2}{I_4} \overline{D}\left(\frac{1}{3}\right), \quad \overline{D}\left(\frac{1}{3}\right) = \lim_{z \to 1/3} \frac{D(z)}{(1-3z)}, \tag{144}$$

as a result, the Gram matrix disappears in expression (117) and a matrix $G$ with the elements

$$G_{ij} = 3 \int d^l x\, h_i(x) \phi_c^2(x) h_j(x), \tag{145}$$

appears instead of $\Gamma$; the dependence of $G$ on the collective variables is the same as for $\Gamma$. As a result, for $d = 4$, we obtain

$$Z_N = \frac{(-S\{\phi_c\})^{-(M+r)/2}}{(2\pi)^{1+r/2}} S\{\phi_c\}^{-N} \Gamma\left(N + \frac{M+r}{2}\right)$$

$$\times \left[ -\overline{D}(1) \overline{D}^{n-1}\left(\frac{1}{3}\right) \right]^{-1/2} I_4^{(n-1)/2} \int d^l u\, \delta(|\mathbf{u}| - 1) u_{\alpha_1} \dots u_{\alpha_M}$$

$$\times \int (d^l x_0)(d \ln \rho)\, d\tau_g (\det \overline{G})^{1/2} \rho^{-4-M}$$

$$\times \overline{\phi}_c\left(\frac{\hat{g}^{-1}(x_1 - x_0)}{\rho}\right) \dots \overline{\phi}_c\left(\frac{\hat{g}^{-1}(x_M - x_0)}{\rho}\right), \tag{146}$$

while for $d < 4$, we must put $\rho = 1$ and eliminate integration with respect to $\ln \rho$. The normalization of determinants is carried out in the conventional way [19, 24] by separating diverging factors and compensating them by their counterterms. As a result, quantity $\overline{D}(1)$ and $\overline{D}(1/3)$ are simply replaced by $\overline{D}_R(1)$ and $\overline{D}_R(1/3)$ for $d < 4$, while for $d = 4$, the substitution

$$[-\overline{D}(1) \overline{D}^{n-1}(1/3)]^{-1/2} \longrightarrow [-\overline{D}_R(1) \overline{D}_R^{n-1}(1/3)]^{-1/2}$$

$$\times \exp(v \ln \mu \rho) \exp\left[ -\frac{3}{4} r + v\left(\frac{1}{2}\ln\frac{4}{3} - \frac{\tilde{I}_4}{I_4}\right) \right], \tag{147}$$

_____________
[19]After the substitution $y = \hat{g}x$, generators $T_i$ are transformed in terms of one another. For $d = 2$ and $d = 3$, the determinant of the transformation is equal to unity and the dependence of $\det \Gamma$ on $\hat{g}$ is absent; apparently, this also holds in the general case.

takes place, where

$$\tilde{I}_4 = \int \frac{d^4 q}{(2\pi)^4} \langle \phi_c^2 \rangle_q^2 \ln q \tag{148}$$

and $\langle \phi_c^2 \rangle_q$ is the Fourier component of function $\phi_c^2(x)$.

To pass to the asymptotic form of $A_K$ (see Section 4), we must carry out the substitution $\phi_c \longrightarrow \psi_c$ and $r \longrightarrow r'$ in all expressions and represent expression (146) in the form (49). Then we have for $A_K$ formula (47) with parameters (50), where $S_1 = S\{\psi_c\}$, $b_1 - b = d(d-1)/4$, and

$$c_1 = \frac{(-S_1)^{(-(M+r'))/2}}{(2\pi)^{1+r'/2}} \left[ \overline{D}_R(1) \overline{D}_R^{n-1}\left(\frac{1}{3}\right) \right]^{-1/2} I_4^{(n-1)/2}$$

$$\times \int d^l u\, \delta(|\mathbf{u}| - 1) u_{\alpha_1} \dots u_{\alpha_M} \int d^l x_0\, d\tau_g (\det \overline{G})^{1/2}$$

$$\times \psi_c(\hat{g}^{-1}(x_1 - x_0)) \dots \psi_c(\hat{g}^{-1}(x_M - x_0)) \tag{149}$$

for $d < 4$ and

$$c_1 = \frac{(-S_1)^{-(M+r')/2}}{(2\pi)^{1+r'/2}} \left[ \overline{D}_R(1) \overline{D}_R^{n-1}\left(\frac{1}{3}\right) \right]^{-1/2}$$

$$\times I_4^{(n-1)/2} \exp\left[ -\frac{3}{4} r + v\left(\frac{1}{2}\ln\frac{4}{3} - \frac{\tilde{I}_4}{I_4}\right) \right]$$

$$\times \int d^l x_0\, d\ln \rho\, d\tau_g (\det \overline{G})^{1/2} \rho^{-4-M} \exp(v \ln \mu \rho)$$

$$\times \psi_c\left(\frac{\hat{g}^{-1}(x_1 - x_0)}{\rho}\right) \dots \psi_c\left(\frac{\hat{g}^{-1}(x_M - x_0)}{\rho}\right) \tag{150}$$

for $d = 4$; here, $r' = r + d(d-1)/2$. All quantities appearing in this formula can be calculated if the form of instanton $\psi_c(x)$ is known.

## ACKNOWLEDGMENTS

This study was financed by the Russian Foundation for Basic Research (grant no. 03-02-17519).

*Translated by N. Wadhwa*